\documentclass[%
 reprint,
nofootinbib,
 amsmath,amssymb,
 aps,
prd,
]{revtex4-2}

\usepackage[normalem]{ulem}
\usepackage[dvipsnames]{xcolor}
\usepackage{graphicx}
\usepackage{dcolumn}
\usepackage{bm}
\usepackage{hyperref}
\usepackage{appendix}
\usepackage{soul}

\newcommand{\aap}{{Astron. Astrophys.}}

\begin{document}

\preprint{APS/123-QED}


\title{DESIgning concordant distances in the age of precision cosmology: the impact of density fluctuations}

\author{David Camarena}
\email{dcamarena93@unm.edu} 
\affiliation{Department of Physics and Astronomy, University of New Mexico, Albuquerque, New Mexico 87106, USA}
\author{Kylar Greene}
\email{kygreene@unm.edu}
\affiliation{Department of Physics and Astronomy, University of New Mexico, Albuquerque, New Mexico 87106, USA}
\author{John Houghteling}
\email{jhoughteling@unm.edu}
\affiliation{Department of Physics and Astronomy, University of New Mexico, Albuquerque, New Mexico 87106, USA}
\author{Francis-Yan Cyr-Racine}
\email{fycr@unm.edu}
\affiliation{Department of Physics and Astronomy, University of New Mexico, Albuquerque, New Mexico 87106, USA}

\date{\today}

\begin{abstract}

Discrepancies between distance measurements and $\Lambda$CDM predictions reveal notable features in the distance–redshift relation, possibly suggesting the presence of an evolving dark energy component. Given the central role of the Friedmann-Lema\^{i}tre-Robertson-Walker (FLRW) metric in modeling cosmological distances, we investigate here whether these features instead point to a possible departure from the fundamental FLRW symmetries. Exploiting the transverse and line-of-sight distances provided by baryonic acoustic oscillations (BAO) observations, we demonstrate that observed distances hint at a slight but systematic preference for an anisotropic expansion rate emerging regardless of the dark energy model considered. Leveraging this non-FLRW feature, we investigate an inhomogeneous extension of the $\Lambda$CDM model that naturally provides an anisotropic expansion rate. Our analysis demonstrates that models featuring spherical overdensities can explain BAO, supernova, and cosmic microwave background data, providing fits statistically indistinguishable from those obtained with a phantom dark energy scenario. When Pantheon+ data is considered, our analysis challenges the FLRW framework at $2.8\sigma$ and yields scenarios that can be interpreted as subtle but non-negligible deviations from the FLRW metric. When DESY5 supernovae are considered instead, deviations are notably more significant, yielding scenarios that mildly violate the Copernican principle and exclude the FLRW assumption at $5.2\sigma$. Overall, our results motivate a more in-depth investigation of whether the perfectly homogeneous and isotropic FLRW paradigm can still be assumed to accurately predict cosmological distances in the era of precision cosmology.

\end{abstract}

\maketitle


\section{Introduction}\label{sec:intro}

Recent probes of the distance-redshift relation, including those based on the baryonic acoustic oscillation (BAO) feature imprinted on the large-scale distribution of galaxies \cite{DESI:2024mwx,DESI:2025zgx} and those based on the apparent luminosity of type Ia supernovae (SNeIa)~\cite{Scolnic:2021amr,Rubin:2025vnz,DES:2024jxu}, might be pointing towards deviations from the standard $\Lambda$ Cold Dark Matter ($\Lambda$CDM) cosmological model. Detailed fitting of those distances appears to necessitate a late-time Hubble expansion history that significantly departs from what is possible within the realm of $\Lambda$CDM. 

When interpreted as a deviation from the standard cosmological constant ($\Lambda$) using the Chevallier-Polarski-Linder (CPL) \cite{Chevallier:2000qy,Linder:2002et} $w_0w_a$ parameterization, the observations seem to favor a time-dependent dark energy component with a phantom equation of state ($w_{\rm eff} < -1$)~\cite{DESI:2025fii}. However, as pointed out in the literature (see e.g.~Refs.~\cite{Cortes:2024lgw,Cortes:2025joz,Wang:2025bkk,Dinda:2025svh,Colgain:2025nzf,Efstathiou:2025tie}), this parameterization suffers from important degeneracies that make it challenging to assess rigorous statistical significance for $\Lambda$CDM deviations. Beyond this specific parameterization, directly measuring the dark energy equation of state (EoS) is difficult due to the indirect dependence of cosmological distances on this quantity \cite{Wang:2004ru,Wang:2025vfb,Nesseris:2025lke,Gonzalez-Fuentes:2025lei}.   
 
Despite these issues, model-independent analyses of cosmological distances \cite{DESI:2025fii,Scherer:2025esj,Ormondroyd:2025iaf,Kessler:2025kju,Specogna:2025guo,You:2025uon,Lee:2025pzo,Cheng:2025lod} still show a marked preference for a dark-energy phantom crossing at redshift $z\sim0.5$ (see however Ref.~\cite{Chen:2025jnr}). Although it is possible to write down physical models that can mimic such dynamics while not violating the null-energy condition (see e.g.~Refs.~\cite{Frion:2023xwq,Wolf:2024eph,Wolf:2025jed,Khoury:2025txd,Chen:2025mlf,Chen:2025wwn,Aoki:2025bmj,Lin:2025gne,deSouza:2025rhv,Dinda:2025iaq,Mirpoorian:2025rfp,Wang:2025zri,Kumar:2025etf,Murai:2025msx,Wolf:2025jed,Akrami:2025zlb,Pan:2025qwy,Chaussidon:2025npr,Silva:2025hxw,Chakraborty:2025syu,Li:2025dwz,Mishra:2025goj,Bhattacharjee:2025xeb,Braglia:2025gdo,Shiu:2025ycw,Bedroya:2025fwh,Philcox:2025faf,Giani:2025hhs}), it is important to consider whether the observed cosmological distances could be explained by a different, perhaps simpler, mechanism. One possible direction is to explore whether all the theoretical assumptions underpinning $\Lambda$CDM are valid in the current era of precision cosmology. Since we are focused here on cosmological distances, a key assumption is that such distances are described by a homogeneous and isotropic Friedmann-Lema\^itre-Robertson-Walker (FLRW) metric. 

While large-scale homogeneity has so far been an excellent approximation in cosmological analyses, the ever-increasing precision of distance-redshift relation measurements might eventually reveal subtle deviations from this key assumption. As they can probe the BAO feature both along the line of sight and in the transverse directions, BAO distance measurements are particularly sensitive to deviations from homogeneity since these two distinct observables are affected differently by large-scale density fluctuations. From our perspective here on Earth, such local inhomogeneities would lead to an apparent anisotropic expansion\footnote{We note that this apparent anisotropic expansion is different than the global anisotropic expansion scenario considered in, e.g.~, Refs.~\cite{Hertzberg:2024uqy,Palacios-Cordoba:2025jmg}.} history in which the line-of-sight and transverse BAO distances no longer follow the FLRW expectation. In principle, this makes even a small breakdown of large-scale homogeneity eminently distinguishable from dynamical dark energy, since the latter typically assumes an underlying FLRW universe.    

Inhomogeneous cosmological models have been proposed to address observational phenomena such as the Hubble tension (see e.g.~Refs.~\cite{Tokutake:2017zqf,Hoscheit:2018nfl,Lukovic:2019ryg,Ding:2019mmw,Kenworthy:2019qwq,Cai:2020tpy,Castello:2021uad,Mazurenko:2023sex,Monjo:2023eby,Martin:2021wvb,Giani:2023aor,Giani:2024nnv,Jia:2025prq,Kraiselburd:2025gti}) and cosmic acceleration~\cite{Celerier:1999hp,Tomita:2001gh,Alnes:2005rw,Biswas:2006ub,Chung:2006xh,Enqvist:2007vb,Garcia-Bellido:2008vdn,Clifton:2008hv,Celerier:2009sv,Clifton:2009kx,February:2009pv,Moss:2010jx,Zhang:2010fa,Zibin:2011ptm,
Bolejko:2011jc,Marra:2011ct,Lapi:2023plb,Lane:2023ndt,Seifert:2024bqr}. Many of these scenarios rely on late-time matter inhomogeneities that significantly violate the Copernican principle—the assumption that we are typical observers in the Universe—or fail to fully account for cosmological observations, raising doubts about their cosmological plausibility~(see e.g.~Refs.~\cite{Camarena:2022iae,Huterer:2023ldv}). However, more modest deviations from the FLRW background remain observationally viable and reveal a nontrivial parameter space encompassing inhomogeneous models that are consistent with, or only mildly violate, the Copernican principle~\cite{Camarena:2021mjr}. Such scenarios feature cosmological distances that notably deviate from the standard predictions of a homogeneous and isotropic Universe, and this deviation may already be testable given the current precision of cosmological distance measurements.

In this paper, we employ the recent data release~2~(DR2) BAO measurements from the Dark Energy Spectroscopic Instrument (DESI), along with SNeIa distances coming from the Pantheon+~\cite{Scolnic:2021amr} and Dark Energy Survey Year 5 (DESY5)~\cite{DES:2024jxu} datasets, to geometrically test the standard homogeneity assumption of the $\Lambda$CDM model. We point out that analyses considering line-of-sight and transverse cosmological distances separately yield slightly different cosmic expansion histories, raising concerns that the Universe is inconsistent with the FLRW expectation (see Refs.~\cite{Teixeira:2025czm,Li:2025htp,Kanodia:2025jqh,Afroz:2025iwo} for related analyses comparing the consistency of luminosity and angular diameter distances). Although current observational uncertainties limit the statistical significance of this feature, we use this observation to motivate a preliminary investigation into whether a small but measurable breakdown of homogeneity could explain the features seen in BAO and SNeIa distance measurements. Using a simple toy model based on the Lema\^itre-Tolman-Bondi (LTB) metric~\cite{Lemaitre:1933gd,Tolman:1934za,Bondi:1947fta}, we demonstrate that a $\Lambda$CDM extension featuring late-time inhomogeneities in the matter distribution could be responsible for the deviation between observed BAO (and SNeIa) distances and the prediction of a $\Lambda$CDM model calibrated to cosmic microwave background (CMB) observations. 

This paper is organized as follows. In Section~\ref{sec:early_hints}, we briefly review how BAO measurements enable simultaneous determination of line-of-sight and transverse BAO distances. We also discuss the consistency between these two distinct measurements, highlighting how they prefer different cosmic expansion histories. In Section~\ref{sec:inhomo_cosmo}, we introduce a simple inhomogeneous extension of the $\Lambda$CDM model and qualitatively explain our modeling choices. Cosmological constraints on this model are presented in Section~\ref{sec:results}, where we also evaluate its goodness of fit. The cosmological implications of such constraints are discussed in Section \ref{sec:disc}. We finally conclude in Section \ref{sec:conc}.

\section{Early hints of an anisotropic expansion}\label{sec:early_hints}

\subsection{Transverse and line-of-sight BAO distances}\label{sec:bao_distances}

The characteristic scale imprinted by BAO in the clustering of matter allows a determination of cosmological distances in two different directions: along, $D_\mathrm{H}$, and transverse, $D_\mathrm{M}$, to the line-of-sight. More specifically, the BAO technique measures the ratios $D_\mathrm{H}/r_\mathrm{d}$ and $D_\mathrm{M}/r_\mathrm{d}$, where $r_\mathrm{d}$ is the comoving drag sound horizon scale. Within the standard paradigm, such distances directly relate to the Hubble rate $H(z)$ via equations:
\begin{equation} \label{eq:DM_FLRW}
    D_\mathrm{M}(z) =  \int_0^z \frac{\mathrm{d}z'}{H(z')}\,,
\end{equation}
and 
\begin{equation} \label{eq:DH_FLRW}
    D_\mathrm{H}(z) = \frac{1}{H(z)} \,,
\end{equation}
where we have assumed zero global curvature and set the speed of light $c=1$.

The importance of BAO measurements in testing cosmological models is difficult to overstate. Not only are they complementary to other probes of cosmological distances---for instance, the ones provided by SNeIa data, whose measurements are also related to $D_\mathrm{M}$\footnote{Standard candles probe the luminosity distance $D_{\rm L}(z)$ through measurements of the distance modulus $\mu(z)$. If gravity is described by a metric theory and photons travel along null geodesics while conserving their number, the Etherington relation holds~\cite{Etherington:1933are}, allowing $D_{\rm L}$ measurements to be directly related to transverse distances via $D_{\rm L}(z) = (1+z) D_{\rm M}(z)$.}---but they also provide a direct test of the expansion history $H(z)$ and of the physics of the early Universe via $r_{\rm d}$. By simultaneously constraining $D_\mathrm{M}(z)$ and $D_\mathrm{H}(z)$ at the same redshift, BAO measurements allow an important test of the fundamental symmetries underpinning our cosmological model, including homogeneity and isotropy. 
 
In the next subsection, we use the DESI DR2 BAO data to test the geometric consistency of the $\Lambda$CDM model, using the standard $w_0w_a$ parametrization as a comparison point. Analyses involving transverse comoving distances, $D_\mathrm{M}$, use BAO distances alone and in combination with SNeIa measurements. On the other hand, constraints that directly probe the Hubble parameter $H(z)$ via $D_\mathrm{H}$ rely solely on BAO data. For the $w_0w_a$ parametrization, we also include analyses that supplement the distance measurements with a compressed version of the latest Planck CMB data~\cite{Planck:2020olo}. This compressed version provides a multivariate Gaussian prior on the parameters $(\theta_*, \omega_{\rm b}, {\omega_{\rm c}})$, enabling constraints on the early-time Universe that are agnostic to the details of the late-time effects~\cite{Lemos:2023xhs}. Here, $\theta_*$ denotes the angular acoustic scale, while $\omega_{\rm b}$ and $\omega_{\rm c}$ represent the physical densities of baryons and dark matter, respectively. Further details of this analysis can be found in Appendix~\ref{sub:ap_geo}. For comparison, we also include some of the constraints reported by DESI~\cite{DESI:2024mwx,DESI:2025zgx} and DESY5~\cite{DES:2024jxu}. 

\subsection{Results of a geometric test: $D_\mathrm{M}$ vs $D_\mathrm{H}$ \label{sub:geo_test}}

\begin{figure}
    \centering
    \includegraphics[width=0.975\linewidth]{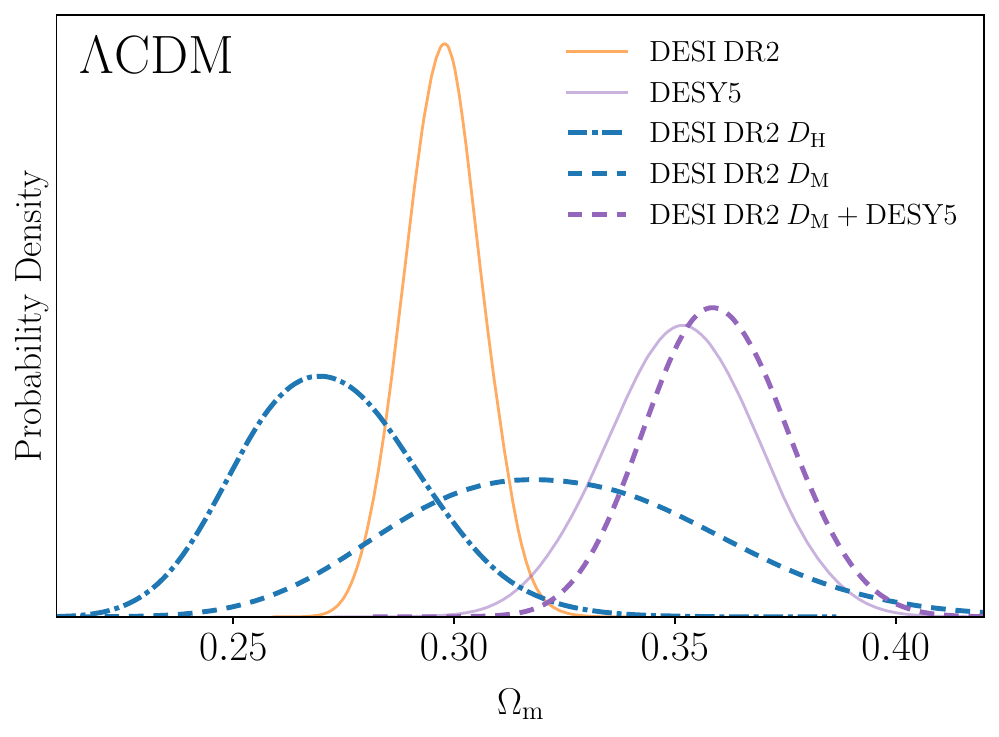}
    \includegraphics[width=0.975\linewidth]{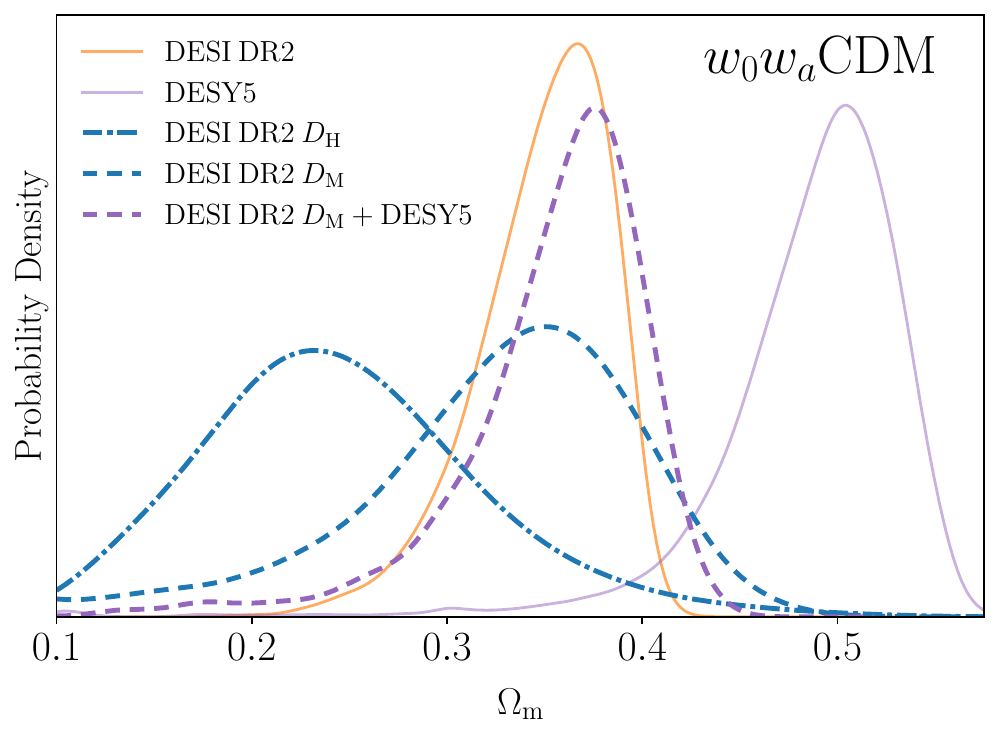}
    \caption{Marginalized $\Omega_\mathrm{m}$ posterior obtained from the geometrical test, $D_\mathrm{M}$ vs. $D_\mathrm{H}$, when considering the $\Lambda$CDM model (upper panel) and $w_0w_a$ parametrization (lower panel).}
    \label{fig:Omega_m}
\end{figure}

The Friedmann equation provides a direct link between the expansion rate $H(z)$ and the energy content of the Universe. As a result, our geometrical tests translate into constraints on the parameters appearing on the right-hand side of the Friedmann equation. In the case of the flat $\Lambda$CDM model, constraints on the late-time expansion rate directly inform the inferred value of the matter density parameter $\Omega_{\rm m}$. For the $w_0w_a$ parametrization, the interpretation is more complex: constraints in the expansion history affect both $\Omega_{\rm m}$ and the dark energy parameters $w_0$ and $w_a$. However, due to statistical limitations and strong correlations between $w_0$ and $w_a$, the impact on these parameters tends to be more moderate compared to that on $\Omega_{\rm m}$. In this section, we present the results of the geometrical test in the context of constraints on $\Omega_{\rm m}$, while the discussion of the effects observed in the $w_0$--$w_a$ parameter space is deferred to Appendix~\ref{ap:geo_w0wa}.

Figure~\ref{fig:Omega_m} shows the marginalized constraints on $\Omega_{\rm m}$ obtained from the geometrical test under the $\Lambda$CDM (upper panel) and $w_0w_a$CDM (lower panel) models. In both cases, we find that cosmological distances measured along the line of sight (dot-dashed blue lines) yield systematically lower values of $\Omega_{\rm m}$ compared to transverse measurements (dashed blue and purple lines). While the differences are generally moderate to small, this trend is nontrivial: within the FLRW framework, both $D_{\rm M}$ and $D_{\rm H}$ are expected to recover the same expansion history, as encoded in Eqs.~\eqref{eq:DM_FLRW} and \eqref{eq:DH_FLRW}. The persistence of this discrepancy across models with different dark energy components may point to a departure from the fundamental assumptions of the FLRW paradigm. In particular, this geometrical feature hints at an anisotropic expansion, whose nature is inconsistent with the assumption of large-scale homogeneity. Furthermore, these results indicate that the $\Omega_{\rm m}$ discrepancy between BAO and SNeIa datasets previously reported in the literature~\cite{DESI:2024mwx,Colgain:2024mtg,Efstathiou:2025tie}
could correspond to a discrepancy between the transverse and line-of-sight expansion rather than a discrepancy between different probes. {Results with SNeIa from Pantheon+, showing similar trends but smaller shifts in $\Omega_{\rm m}$, are presented in Appendix~\ref{ap:geo_Pantheon}.}

Overall, Figure~\ref{fig:Omega_m} points out that slightly different expansion histories are obtained when considering the line-of-sight and transverse distances separately. 
Although the statistical significance of this feature is limited, it is important to acknowledge that the overall trend, especially the one observed in $\Omega_\mathrm{m}$, is rather non-trivial as the feature emerges regardless of the nature of the dark energy component. Results of our geometrical analysis raise concerns about whether the observed cosmological distances can be accurately modeled within the FLRW paradigm. 
This also provides features distinguishable from dynamical dark energy scenarios and motivates the study of beyond-FLRW scenarios capable of explaining the observed cosmological distances. 
One of such scenario is given by inhomogeneous extensions of $\Lambda$CDM, which despite being tightly constrained by current data~(see e.g. Ref.~\cite{Camarena:2021mjr}), can still feature different distance-redshift relations and simultaneously provide a natural explanation for an anisotropic expansion rate. We show below how a simple inhomogeneous generalization of the $\Lambda$CDM model can achieve this.

\section{Inhomogeneous cosmology}\label{sec:inhomo_cosmo}

\subsection{Motivation for inhomogeneous models}

Although large-scale homogeneity has been foundational in cosmology, it is ultimately an approximation rather than a fundamental principle.
The FLRW metric inherently represents a coarse-grained description, averaging out smaller-scale structures on the order of few hundreds Mpc.
Yet, as observational precision improves, subtle effects from local density fluctuations at intermediate scales ($0.01 \lesssim z \lesssim 0.1$) become increasingly more relevant.
Due to their gravitational influence, these intermediate-scale perturbations might alter the local expansion rates---and, thus, the distance-redshift relation---in ways not fully captured by the FLRW metric.

The discrepancies in $\Omega_{\rm m}$ highlighted in the previous section and the tension between the observed distances and the predictions of the $\Lambda$CDM model \cite{DESI:2025zgx} could reflect subtle yet significant variations in the local expansion histories of different patches of the Universe (see also Ref.~\cite{Gialamas:2024lyw}). Intermediate-scale inhomogeneities perturbing the metric away from the FLRW case could potentially reconcile the observed distance measurements without invoking exotic dark-energy behavior.
Indeed, density fluctuations could yield an apparent evolution of the dark energy EoS (see, e.g., Ref.~\cite{Romano:2010nc,Valkenburg:2013qwa}).
Compared to a phantom dark-energy component, inhomogeneous scenarios would admit a more straightforward physical interpretation: due to its assumption of homogeneity over a wide redshift range, the $\Lambda$CDM model falls short in describing the observed cosmological distances, particularly those at local and intermediate scales. 

To test whether the current distance discrepancies observed in cosmological data might simply reflect local inhomogeneities rather than a departure from the cosmological constant, we employ a toy model featuring a large-scale radial inhomogeneity as a diagnostic framework. We model such a scenario using the LTB metric with a non-vanishing cosmological constant $\Lambda$. We emphasize that although these models, usually quoted as $\Lambda$LTB models (see e.g Ref.~\cite{Marra:2011ct,Valkenburg:2012td,Camarena:2021mjr}), do not correspond to the definitive physical description of our local environment, they provide a valuable tool for assessing the robustness of cosmological inferences derived from the assumption of homogeneity. 

The $\Lambda$LTB model admits a wide range of variations over the FLRW metric, allowing us in principle to test a broad range of inhomogeneous models that may or may not be consistent with the assumption of the Copernican principle or the standard inflationary paradigm. In particular, the LTB degrees of freedom can be set up to achieve a rather simple late-time inhomogeneous extension of the $\Lambda$CDM model, whose relative simplicity makes readily available a framework to compute observables that might otherwise be difficult to predict, for instance, the primary observables of the CMB~(see e.g. Ref.~\cite{Clarkson:2007yp,Zibin:2008vj}). In the following, we follow this simple extension. 

Importantly, inhomogeneous cosmologies should not be viewed as radical departures from the standard FLRW paradigm, but rather as a well-motivated extension. 
The FLRW metric itself arises as a special solution to Einstein equations where spatial gradients in matter and curvature are neglected.
By reintroducing a radial dependence into the metric, while retaining spherical symmetry, the LTB metric serves as a relativistic framework that aims to effectively mimic the impact of more complex inhomogeneous structures.

\subsection{Distance measures in the LTB metric}

The standard FLRW metric is built on the foundational assumptions of homogeneity and isotropy. As a result, the expansion of the Universe is governed by a single, spatially uniform Hubble rate, $H(z)$, describing the background evolution. In contrast, inhomogeneous cosmological models challenge this framework, offering a more complex picture of the dynamics of the Universe that allows the expansion rate to vary across space.

For instance, relaxing the assumption of homogeneity while retaining isotropy around a privileged central observer leads to the LTB metric, whose line element generalizes FLRW as~\cite{Lemaitre:1933gd,Tolman:1934za,Bondi:1947fta}:
\begin{equation} \label{eq:LTBmetric}
ds^2 = -dt^2 + X^2(r,t)\,dr^2 + A^2(r,t)\,d\Omega^2,
\end{equation}
where $X(r,t)$ and $A(r,t)$ encode the radial and transverse scale factors, generalizing the single global scale factor of FLRW cosmology\footnote{As a consistency check, the FLRW limit is recovered when
\begin{equation}
A(r,t) = a(t) r, \quad X(r,t) = \frac{a(t)}{\sqrt{1 - k r^2}}.
\end{equation}}.
These functions naturally give rise to two distinct expansion rates: one transverse and one radial,
\begin{equation} \label{eq:LTB_expansions}
    H_{\perp}(r,t) = \frac{\partial_t A(r,t)}{A(r,t)} \quad \text{and} \quad H_{\parallel}(r,t) = \frac{\partial_t X(r,t)}{X(r,t)}\,,
\end{equation}
revealing an intrinsically anisotropic expansion history. Thus, unlike the FLRW metric, in an inhomogeneous LTB metric different cosmological distances---measured along different directions---are governed by different expansion rates, providing a natural explanation to the results of the geometrical test presented in Section~\ref{sub:geo_test}.

In this metric, the transverse comoving distance is given by
\begin{equation}\label{eq:DM_LTB}
D_{\rm M}(z) = (1+z)\; A\left(r(z), t(z)\right),
\end{equation}
where the function $A(r,t)$ physically represents the angular radius of a spherical shell centered on the observer. The line-of-sight Hubble distance follows
\begin{equation}\label{eq:DH_LTB}
    D_{\rm H}(z) = \frac{1}{H_{\parallel}(r(z),t(z))}\,.
\end{equation}
The distance–redshift relation is then fully determined by the geodesic equations, which are given by:
\begin{align}
    \frac{dt}{dz} & = - \frac{1}{(1+z)\, H_{\parallel}(r,t)}\,, \\
    \frac{dr}{dz} & = \frac{1}{(1+z)\, \partial_t X(r,t)}\,.
\end{align}

The impact of density fluctuations on cosmological distances can be schematically understood by examining their effect on the expansion rates. In regions where the local matter density exceeds the cosmic average, gravitational attraction slows down the expansion along the line of sight, leading to a larger $D_{\rm H}$ compared to an equivalent FLRW scenario. Conversely, underdense regions experience a faster local expansion, which increases the line-of-sight Hubble parameter and thus reduces $D_{\rm H}$. A similar, though typically milder, effect arises for the transverse comoving distance $D_{\rm M}$. Owing to the assumption of spherical symmetry, transverse distances are less sensitive to radial density fluctuations, but still respond to the overall distribution of matter through their influence on the transverse expansion rate. As shown in Eq.~\eqref{eq:friedmann_LTB}, the inhomogeneous counterpart of the Friedmann equation, the transverse Hubble rate $H_\perp$, depends on the average matter density.

This anisotropic expansion history has direct consequences for the interpretation of cosmological observables. The most straightforward case involves standard candles, such as SNeIa, whose luminosity distances are given by
\begin{equation}
    D_{\rm L}(z) = (1+z)^2\, A(r(z), t(z)\big)\,,
\end{equation}
when the Etherington relation is assumed. The consequences for BAO observables are instead more intricate. 

The BAO technique does not measure absolute distances directly, but rather the ratio of the imprinted BAO scales to the comoving sound horizon $r_{\rm d}$. We typically express such ratios using comoving distances, since in FLRW cosmologies the conversion between comoving and physical distances is uniform and trivially cancels out. In inhomogeneous cosmologies, however, the mapping between comoving and physical distances depends on time and radial scale, and the conversion factors do not trivially cancel out. Taking into account the induced directional dependencies in the $\Lambda$LTB model, the BAO observed ratios follow (see e.g. Ref.~\cite{Garcia-Bellido:2008xmz,Biswas:2010xm,Marra:2010pg})
\begin{align}
    \frac{D_\mathrm{H}}{r_{\rm d}} & = \frac{1}{l_\parallel(z)(1+z)H_\parallel(r(z),t(z))} \,,\\
    \frac{D_\mathrm{M}}{r_{\rm d}} & = \frac{A(r(z),t(z))}{l_\perp(z)} \,,
\end{align}
where $l_\parallel$ and $l_\perp$ are the physical sound horizon distance at the drag epoch time, $t_{\rm d}$, along the line-of-sight and transverse directions. They are defined as
\begin{align}
    l_\parallel(z) & = \frac{X(r(z),t(z))}{X(r(z),t_{\rm d})}\, \frac{r_{\rm d}}{(1+z_{\rm d})} \,,\\
    l_\perp(z) & = \frac{A(r(z),t(z))}{A(r(z),t_{\rm d})}\, \frac{r_{\rm d}}{(1+z_{\rm d})} \,,
\end{align}
with $z_{\rm d}$ being the redshift corresponding to the drag epoch. 

\subsection{Description of our $\Lambda$LTB model} \label{sub:LLTB_model}

Relaxing the assumption of homogeneity introduces additional degrees of freedom that are absent in the FLRW paradigm. In the $\Lambda$LTB framework, these new degrees of freedom appear as three arbitrary functions, each encoding a different aspect of the space-time geometry. In Appendix~\ref{ap:LTB_extra}, we present and discuss these functions in detail, along with the specific choices adopted in our analysis. 

In summary, our setup defines an extension of the $\Lambda$CDM model, which endows a compensated spherical inhomogeneity characterized by its present-day central matter density contrast, $\delta_0$, and its boundary redshift, $z_{\rm B}$, at which the metric becomes effectively homogeneous. The size of these density fluctuations can be also denoted in terms of the boundary radius, $r_{\rm B}$, which represents the comoving distance between the center to the edge of the spherical inhomogeneity in the FLRW coordinates. 
The particular choice of a compensated profile additionally introduces another characteristic scale, $r_{\rm L}$, which defines the comoving size of the internal overdense or underdense region of the inhomogeneity. Broadly speaking our choices establish $r_{\rm L} \approx \frac{1}{2} \:r_{\rm B}$. Moreover, the impact on the CMB is overall expected to be minimal, as the model is constructed to follow FLRW dynamics at early times. A detailed discussion of how the small changes in the CMB constrain the parameters $\delta_0$ and $z_{\rm B}$ is provided in Appendix~\ref{ap:inho_CMB}.

\section{Constraints on an inhomogeneous Universe}\label{sec:results}

Using a profile likelihood analysis, we investigate how current observations constrain the parameter space of the $\Lambda$LTB model. This is carried out in both one dimension---running over $\delta_0$---and two dimensions, using both $\Lambda$LTB parameters  $(\delta_0, z_{\rm B})$. 
Compared to a full Bayesian exploration, the frequentist approach based on the profile likelihood offers a straightforward and computationally efficient way to preliminarily explore complex, non-Gaussian parameter spaces, without the immediate need for tailored sampling techniques aimed at overcoming the well-known inefficiencies of standard Metropolis–Hastings algorithms. Moreover, since only small correlations are expected between the inhomogeneous parameters and their counterparts in standard $\Lambda$CDM cosmology (see, e.g., Ref.~\cite{Camarena:2021mjr}), the results obtained from the profile likelihood are expected to provide a reliable approximation to those derived from a full Bayesian analysis.

Our analysis considers various combinations of data, including BAO measurements from DESI DR2, Type Ia supernova distances from Pantheon+ and DESY5, and the CMB priors on $(\theta_*, \omega_{\rm b}, \omega_{\rm c})$ described in Section~\ref{sub:geo_test}. While our primary focus is on cosmological distance probes, we include CMB constraints to ensure that the models under consideration not only provide a good fit to the late-time data but also remain consistent with the early-time successes of the $\Lambda$CDM framework. 

Our analysis also quantifies how well observations can be fitted when relaxing the assumption of large-scale homogeneity by comparing the value of $\chi^2$-minimum of the $\Lambda$LTB model with the one obtained when considering the standard $\Lambda$CDM paradigm. We denote such differences as $\Delta \chi^2_{\rm min}$. For comparison, we also present the $\Delta \chi^2_{\rm min}$ obtained when considering an evolving dark-energy component, modeled through the $w_0w_a$ parametrization. Since both the $w_0w_a$ and $\Lambda$LTB scenarios introduce two additional parameters---used to characterize a free function, namely dark energy EoS and curvature profile, respectively---a direct comparison between their respective values of $\Delta \chi^2_{\rm min}$ is possible. We denote the combination of the CMB priors and DESI BAO distances as ``Base''. Further details on the methods employed by this analysis are discussed in Appendix~\ref{sub:profile_like}.

\begin{figure*}
    \centering
    \includegraphics[width=0.535\linewidth]{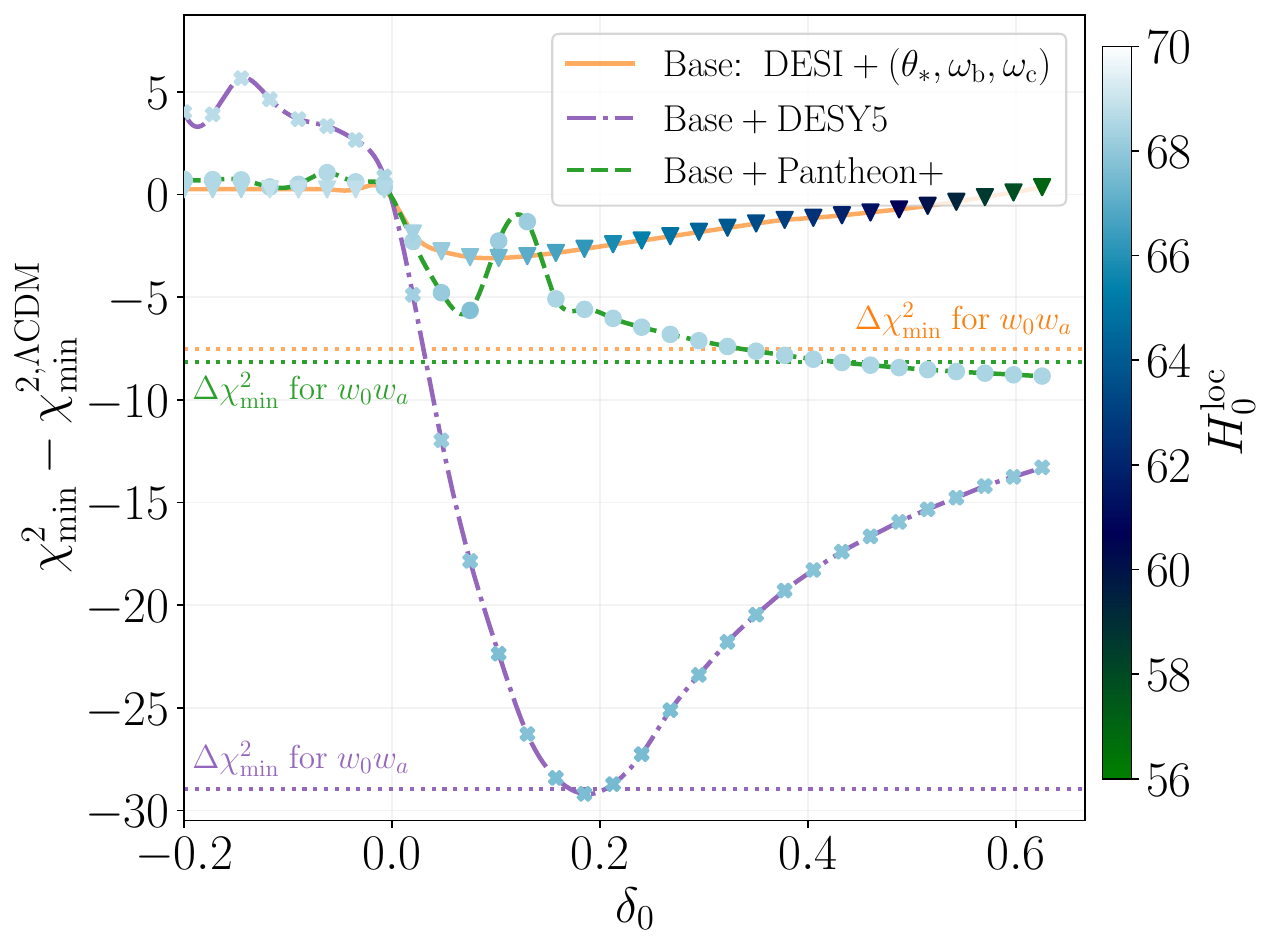}
    \includegraphics[width=0.455\linewidth
    ]{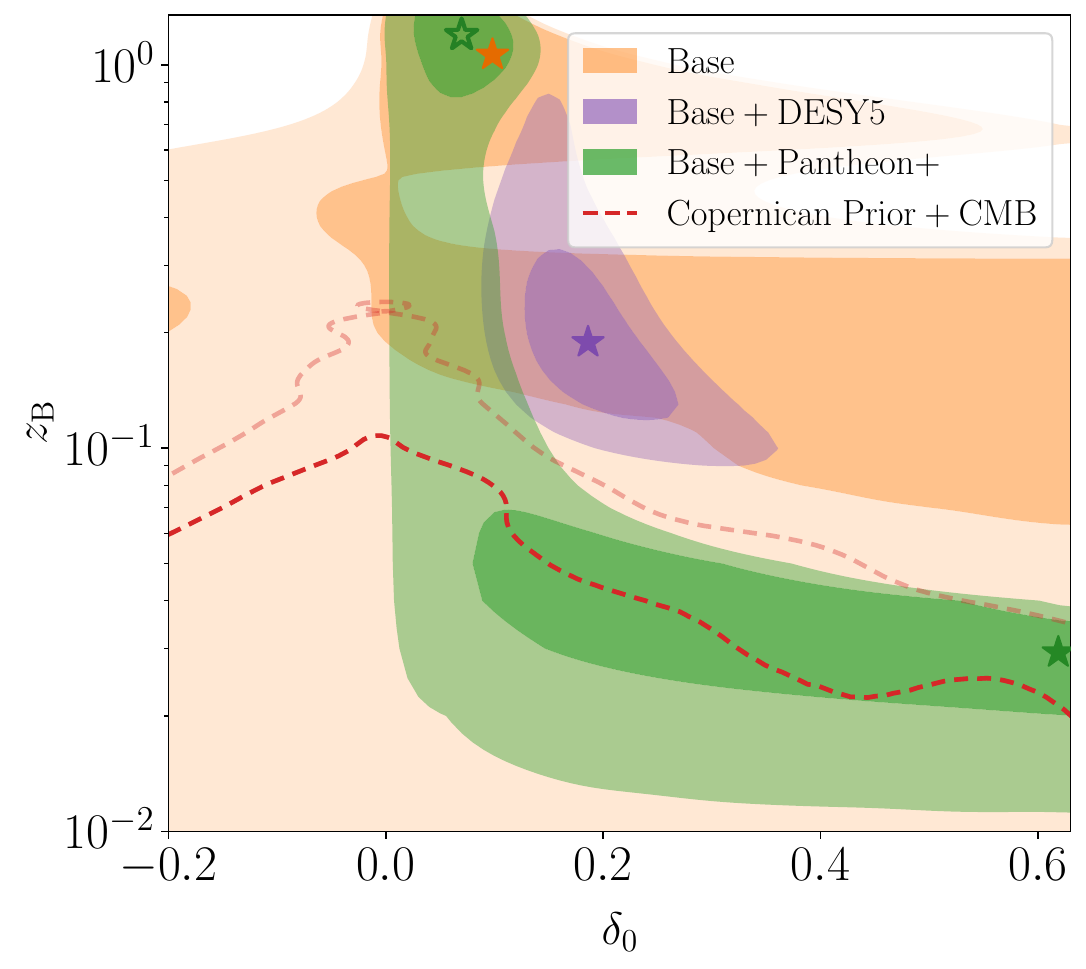}
    \caption{Constraints on the $\Lambda$LTB model obtained by profiling the likelihood in one (left) and two dimensions (right). In the left panel, horizontal dotted lines---color-coded by data combination---indicate the $\chi^2_\mathrm{min}$ values for the $w_0w_a$CDM model relative to the $\Lambda$CDM best fit. In the right panel, filled contours represent the (frequentist) 68$\%$ and 95$\%$ confidence regions for the different data combinations. The dark and light dashed red lines denote the (Bayesian) 68$\%$ and 95$\%$ credible regions, respectively, obtained in Ref.~\cite{Camarena:2021mjr} by applying a Copernican prior convolved with CMB observations.}
    \label{fig:chi2_and_LLTB_contours}
\end{figure*}

\subsection{Dynamic dark energy vs inhomogeneities} 

Results of our profile likelihood in one-dimensional parameter space are shown in the left panel of Fig~\ref{fig:chi2_and_LLTB_contours}, which also displays the value of $\Delta \chi^2_{\rm min}$ obtained for the $w_0w_a$ parametrization (color-coded dotted lines). The left panel furthermore shows how the local expansion rate changes across the parameter space; this change and its cosmological implications are later discussed in Section~\ref{sub:local_H0}.

The left panel of Fig.~\ref{fig:chi2_and_LLTB_contours} reveals that when considering the DESI DR2 data and CMB priors (orange solid line), the $\Lambda$LTB model marginally improves on the fit of $\Lambda$CDM, providing only $\Delta\chi^2_{\rm min} \approx -3.1$. In this case, the inhomogeneous scenario is surpassed by the $w_0w_a$ parametrization, whose fit yields $\Delta\chi^2_{\rm min} \approx -7.5$.

\begin{table}
\caption{\label{tab:chi2_mins} Compilation of the $\Delta{\chi}^2$ values corresponding to $\chi^2$-minimum of the $w_0w_a$CDM (upper table) and $\Lambda$LTB (lower table) models compared to the $\Lambda$CDM best fit when considering different data combinations. Values in parentheses in the last column of the lower table refer to the local minimum found when considering Pantheon+ supernovae.}
\renewcommand{\arraystretch}{1.45}
\begin{ruledtabular}
\begin{tabular}{cccc}
  & Base & Base + DESY5 & Base + Pantheon+ \\ \hline \hline
\multicolumn{4}{c}{$w_0w_a$CDM} \\ \hline
 $\Delta{\chi}^2_\mathrm{BAO}$ & -4.18 & -5.57 & -2.72 \\
 $\Delta{\chi}^2_\mathrm{CMB}$ & -3.36 & -1.66 & -1.96 \\
 $\Delta{\chi}^2_\mathrm{SNe}$ & - & -21.73 & -3.47 \\ \hline
 $\Delta {\chi}^2_\mathrm{min}$ & -7.54 & -28.96 &  -8.15 \\ \hline \hline
\multicolumn{4}{c}{$\Lambda$LTB} \\ \hline
 $\Delta{\chi}^2_\mathrm{BAO}$ & -3.50 & -2.18 & -0.04 (-4.07) \\
 $\Delta{\chi}^2_\mathrm{CMB}$ & 0.41 & 1.45 & 0.25 (0.94) \\
 $\Delta{\chi}^2_\mathrm{SNe}$ &  -  & -28.5 & -9.05 (-2.49) \\ \hline
 $\Delta {\chi}^2_\mathrm{min}$ & -3.09 & -29.21 & -8.84 (-5.64) 
\end{tabular}
\end{ruledtabular}
\end{table}

With the addition of SNeIa data, however, the results become notably more interesting. The analysis with Pantheon+ (dashed green line) found that the $\Lambda$LTB model fits the data significantly better than the $\Lambda$CDM scenario, attaining $\Delta \chi^2_{\rm min} \approx -8.8$. Remarkably, this fit is statistically indistinguishable from the fit provided by the $w_0w_a$ parametrization, which yields $\Delta \chi^2_{\rm min} \approx -8.2$. A similar trend is found when using DESY5 supernovae instead (dot-dashed purple line); the $\chi^2$-minimum values for the $\Lambda$LTB and $w_0w_a$ cases are nearly identical and both provide a significantly better fit to the data than the standard $\Lambda$CDM framework, specifically, $\Delta \chi^2_{\rm min} \approx -29$. It is additionally notable that the profile likelihood obtained when considering Pantheon+ is bimodal, with a local minimum that features $\Delta\chi^2_{\rm min} \approx -5.6$ and it is similar to the best fit found without supernovae. 

Table~\ref{tab:chi2_mins} provides a detailed breakdown of the goodness of fit for the $\Lambda$LTB and $w_0w_a$CDM models, decomposing the total $\Delta \chi^2_{\rm min}$ into contributions from each cosmological probe. Consistent with the assumption of an inhomogeneous late-time generalization of the standard paradigm (see Section~\ref{sub:LLTB_model} and Appendix~\ref{ap:LTB_extra}), the $\Lambda$LTB and $\Lambda$CDM models provide statistically equivalent fits to the CMB data across all combinations, with the largest deviation being $\Delta \chi^2_{\rm CMB} \approx 1.5$. When supernova data are excluded, the inhomogeneous model provides a modest improvement in fitting BAO distances, resulting in a $\Delta \chi^2_{\rm BAO}$ only slightly larger than that achieved by the $w_0w_a$ parametrization. However, the inclusion of supernovae changes this trend. In such cases, the $\Lambda$LTB trades the improvement on BAO fitting  in exchange for a significantly improved fit to the SNeIa data. Compared to the $w_0w_a$ framework, $\Lambda$LTB provides a better fit to the supernova data but performs slightly worse on BAO distances. These changes on the SNeIa and BAO fits compensate for each other, ultimately yielding a total $\Delta \chi^2_{\rm min}$ that closely matches the $w_0w_a$ fits. 

The profile likelihood analysis in $\delta_0$ additionally reveals a systematic preference for a spherically overdense region, $\delta_0 > 0$, whose amplitude and size vary across different data combinations. This is also illustrated in the right panel of Fig.~\ref{fig:chi2_and_LLTB_contours}, which presents the results of the two-dimensional profile likelihood analysis, along with the corresponding best-fit points (color-coded stars).

The two-dimensional profile likelihood shows that the preference for  $\delta_0 > 0$ is small when considering the BAO distances calibrated by CMB priors (orange contours). In this case, FLRW limit remains allowed at approximately $1.5\sigma$ and the best fit (orange star) corresponds to an overdensity with $\sim 10\%$ more matter at its center than the global average and a significantly large comoving radius of $r_{\rm B} = 3.5\;{\rm Gpc}$ ($z_{\rm B} \approx 1.1$).

\begin{figure*}
    \centering
    \includegraphics[width=0.995\linewidth]{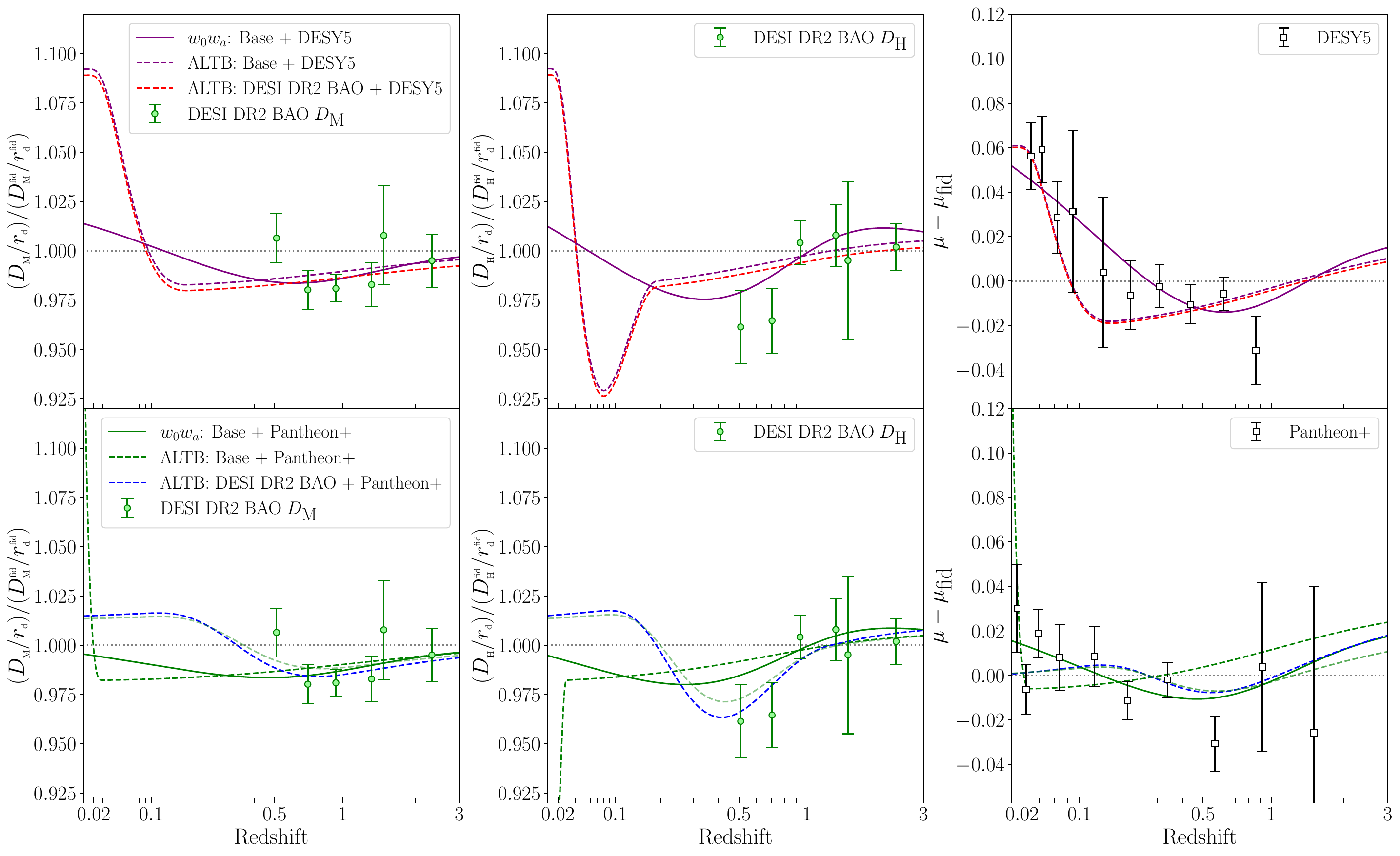}
    \caption{The transverse (left column) and line-of-sight (middle column) BAO distances, along with the SNe~Ia distance modulus (right column), predicted for the best-fit $w_0w_a$ and $\Lambda$LTB models. These are compared to the fiducial $\Lambda$CDM cosmology used by DESI in their BAO reconstruction analysis. Results are presented separately for each supernova dataset, with upper and lower panels corresponding to DESY5 and Pantheon+, respectively. The light green dashed line in the lower panels denotes the secondary local minimum identified in the Pantheon+ analysis. For clarity, the SNeIa data are shown in binned form; the binning procedure is described in Appendix~\ref{sub:SNe_binning}.}
    \label{fig:6_panel}
\end{figure*}

The analysis incorporating Pantheon+ (green contours), on the other hand, allows for the assumption of large-scale homogeneity at roughly $2.8\sigma$, with the best fit model (solid green star) featuring a significantly smaller and denser region: $r_{\rm B} \approx 130\;{\rm Mpc}$ ($z_{\rm B} \approx 0.03$) and $\delta_0 \approx 0.6$. The local minimum arising in this case (empty green star) closely resembles the best fit obtained without supernova data. When using DESY5 observations instead, the preference for an overdense region is further amplified, leading to exclude the FLRW case at around $5.4\sigma$. In comparison to Pantheon+ supernovae, the best fit to DESY5 distances leads to a larger but shallower inhomogeneity, represented by $r_{\rm B} \approx 800 \;{\rm Mpc}$ ($z_{\rm B} \approx 0.19$) and $\delta_0 \approx 0.2$. Notably, similar to the standard analysis of BAO distances within the $w_0w_a$ framework~\cite{DESI:2025zgx}, the DESY5 distances yield models with more remarkable departures from the predictions of the $\Lambda$CDM model than the Pantheon+ supernovae. {While the source of the difference between DESY5 and Pantheon+ merits further investigation---an issue already discussed in previous studies~\cite{Notari:2024zmi,Gialamas:2024lyw,Efstathiou:2025tie}---a detailed analysis is beyond the scope of this work.}

Fig.~\ref{fig:6_panel} shows the distance-redshift relations predicted by the best-fit overdensities and the $w_0w_a$ best-fit models in comparison to the fiducial $\Lambda$CDM model used in the DESI BAO analysis~\cite{DESI:2024cru}. As previously discussed, when supernovae are included, the best-fit models significantly improve their fit to the SNeIa data at the cost of worsening the fit to BAO distances. Interestingly, this leads to models with notable features at low redshifts. 

Our analysis explicitly demonstrates that cosmological distances can be consistently explained by modifications to the distance–redshift relation induced by density fluctuations. In most cases, such scenarios yield fits comparable to the widely adopted $w_0w_a$ parametrization. However, achieving a competitive fit---particularly for models constrained by DESY5 measurements---requires matter density fluctuations extending to Gpc scales, raising concerns about their cosmological plausibility within the observable Universe. In Section~\ref{sub:rare}, we discuss the likelihood of finding these best-fit overdensities under the assumptions of the Copernican principle and a nearly scale-invariant primordial curvature power spectrum.

\section{Discussion}\label{sec:disc}

\subsection{The local expansion rate}\label{sub:local_H0}

As previously discussed, in regions where the local matter density exceeds the cosmic average, the expansion rate is expected to slow down due to the enhanced gravitational attraction. Consequently, the preference for an overdense region revealed by the profile likelihood analyses raises the concern that such scenarios could exacerbate the Hubble tension. While a Bayesian analysis providing the posterior distribution of the Hubble constant would be required to rigorously assess this impact, we provide a preliminary evaluation by examining how $H_0$ varies across the profile likelihoods under the different data combinations.

Given that the Hubble constant is not well-defined within the $\Lambda$LTB framework, we examine the impact of the observed overdensities on the Hubble tension by adopting a proxy quantity, $H_0^{\rm loc}$. This represents the effective Hubble constant that a central observer would infer using the typical cosmic distance ladder method (e.g., the one adopted by Ref.~\cite{Riess:2021jrx}), while residing inside a spherically symmetric inhomogeneity. This quantity is computed following the procedure outlined in Section~2.3.2 of Ref.~\cite{Camarena:2022iae} (see also Ref.~\cite{Redlich:2014gga}). The variations on $H_0^{\rm loc}$ across the parameter space are indicated by color-coded inverted triangles, circles, and cross marks in the left panel of Fig.~\ref{fig:chi2_and_LLTB_contours}. 

When only BAO distances calibrated by CMB priors are considered (inverted triangle markers), $H_0^{\rm loc}$ varies significantly across the $\Lambda$LTB parameter space, spanning a range of $57$--$68.9\;{\rm km\;s}^{-1}\;{\rm Mpc}^{-1}$. However, this variation is substantially reduced with the inclusion of SNeIa data (circle and cross markers), which constrain $H_0^{\rm loc}$ to a narrower range of approximately $67.5$--$68.8\;{\rm km\;s}^{-1}\;{\rm Mpc}^{-1}$. These values, which are consistent with $\Lambda$CDM constraints derived from CMB observations, suggest that the $\Lambda$LTB overdensity is unlikely to exacerbate the Hubble tension. In fact, relative to the $w_0w_a$ parametrization, the inhomogeneous model could slightly alleviate the tension, as the inferred values of $H_0^{\rm loc}$ tend to be higher than the mean values reported by the DESI collaboration in their $w_0w_a$ analysis~\cite{DESI:2025zgx}.

\subsection{How rare are such structures?}\label{sub:rare}

The constraints obtained in the previous section span a wide range of scenarios, from subtle but non-negligible deviations from the FLRW metric on small scales to extreme cases in which homogeneity is not restored until Gpc scales. While these scenarios should not be taken at face value---since the $\Lambda$LTB framework serves as a simplified, exact solution to the Einstein equations intended to approximate the effects of more general inhomogeneous degrees of freedom---it remains essential to assess how rare and plausible such structures are in light of other foundational assumptions underlying the standard paradigm.

For instance, under the assumption of the Copernican principle, the rarity of $\Lambda$LTB structures may be quantified by considering the standard deviation of matter density fluctuations at a given scale, as determined by the matter power spectrum constrained by CMB observations. Specifically, the Copernican prior~\cite{Hunt:2008wp,Valkenburg:2012td} provides a way to estimate the likelihood of realizing a particular inhomogeneity---characterized by specific values of $\delta_0$ and $z_B$---given the matter power spectrum inferred via CMB analysis. Constraints obtained with this prior can then be compared to data constraints, allowing to identify regions of the parameter space that, despite deviating from the FLRW metric, are still consistent with the Copernican principle and the amplitude fluctuations inferred from the CMB power spectrum.\footnote{Although the Copernican principle is often invoked to justify the FLRW assumption~\cite{Ellis:2006fy}, it is important to note that this does not strictly require perfect large-scale homogeneity. In fact, the combination of the Copernican principle with the observed near-isotropy of the CMB leads to an almost-FLRW geometry~\cite{Stoeger:1995egs}, which still allows inhomogeneous deviations consistent with this foundational assumption (see e.g. Ref.~\cite{Clarkson:2012bg}).}

We use the Copernican prior constraints obtained in Ref.~\cite{Camarena:2021mjr} and compare them to the results of our profile likelihood analysis. Such constraints are shown by the dark and light red dashed lines in the right panel of Fig.~\ref{fig:chi2_and_LLTB_contours}. The comparison reveals that analyses excluding SNeIa data yield a broad parameter space, spanning from regions fully consistent with the Copernican principle to those in strong violation with it. Notably, the best-fit model in this case corresponds to a violation of the Copernican assumption at $\sim 100\sigma$, implying an extremely low probability of realizing such an inhomogeneity.

When including Pantheon+ supernovae data, the analysis shows that most of the $\Lambda$LTB scenarios allowed by the data---including the best-fit model---corresponds to subtle deviations from the FLRW metric at intermediate and small scales, specifically, $z_{\rm B} \lesssim 0.25$. Interestingly, 
despite introducing non-trivial modifications to the distance–redshift relation, such scenarios remain consistent with the Copernican prior. Similar to the analysis without supernovae data, models in the parameter space vicinity of the local minimum are extremely non-Copernican. 

When DESY5 observations are used instead, the inhomogeneous models that provide a good fit to the data tend to (moderately) violate the Copernican principle. Although the best-fit model in this case corresponds to a $\sim 10\sigma$ violation, this dataset still allows regions that are only mildly inconsistent with the assumption that we do not occupy a special place in the Universe.

It is crucial to emphasize that the interpretation of the Copernican prior used in Ref.~\cite{Camarena:2021mjr} is bound to the assumption of a nearly scale-invariant primordial curvature power spectrum---consistent with the standard paradigm of inflation
(and some of our particular choices for the LTB metric, see Appendix~\ref{ap:LTB_extra}). Modifications to this assumption, particularly at large scales, could non-trivially alter the inferred probability of finding different structures. 

Nonetheless, modifying this assumption to generate similar large-scale inhomogeneities is a challenging task. Features in the inflationary potential, such as inflection points or plateaus inducing transient ultra-slow-roll phases, can significantly enhance the variance of primordial perturbations at selected wavelengths\cite{Griffiths:2000jb,Leach:2001zf,Inoue:2001zt,Kinney:2005vj}. 
Similarly, multi-field inflation models can naturally produce localized departures from scale invariance \cite{Braglia:2020fms}, and non-canonical inflationary models allowing perturbation propagation speeds to differ from the speed of light \cite{Gariazzo:2016blm,Zhai:2022mpi}, also boosting large-scale power. 
However, such modifications are strongly constrained by current CMB observations, which show little evidence of excess power at horizon scales, limiting their viability in consistently explaining observed Gpc-scale inhomogeneities \cite{Braglia:2020fms}.
A detailed analysis of whether CMB observations permit such modifications---and thereby allow for a broader Copernican region in the parameter space---is left for future work.

\subsection{Beyond a simple $\Lambda$LTB approach}

The arbitrary degrees of freedom of the LTB metric have been configured to produce scenarios that modify the distance–redshift relation at late times, while at the same time preserving the standard model’s success in explaining early Universe phenomena (see Appendix~\ref{ap:LTB_extra}). In particular, our $\Lambda$LTB models assumes
a compensated curvature profile given Eq.~\eqref{eq:kr}~and~\eqref{eq:kr_P}. While the compensating feature is mainly imposed to simplify the approximation to the CMB data, the detailed functional form of this profile is arbitrary. 

Importantly, the fit to the data can vary depending on the assumptions made about the curvature profile. In particular, it may be possible to obtain scenarios that provide a better fit to the data than the $w_0w_a$ parametrization. For example, by modifying the smoothness of the transition between the inner and outer regions of the $\Lambda$LTB inhomogeneity\footnote{Precisely, by considering \begin{align}
P_n(x) = 
\begin{cases}
1  & \text{for } 0 \leq x < \frac{3}{4} \,,\\
1 - \exp\left[-\dfrac{ 4^n (1 - x)^n}{4x-3}\right] & \text{for } \frac{3}{4} \leq x < 1 \,,\\
0 & \text{for } x \geq 1\,,
\end{cases}
\end{align}
instead of Eq.~\eqref{eq:kr_P}.
}, the fit to the Pantheon+ supernovae data can be improved, yielding $\Delta \chi^2_{\rm min} \approx -11.1$, which corresponds to a moderately better fit than the $w_0w_a$ parametrization. In this case, the best-fit overdense region features a more conservative central matter density contrast of $\delta_0 \approx 0.3$ and a smaller boundary radius of $r_{\rm B} \approx 80$ Mpc ($z_{\rm B} \approx 0.018$). This further suggests that adopting different curvature profiles could lead to $\Lambda$LTB scenarios that remain broadly consistent with the Copernican principle.

Our preliminary analysis, based on the $\Lambda$LTB model, overall establish that cosmological distances might simply reflect deviations from the homogeneity assumption rather than a departure from the cosmological constant. 
However, this conclusion is limited by the fact that, due to its simplicity, the $\Lambda$LTB model is unlikely to provide an accurate physical description of our position in the Universe. More complex inhomogeneous degrees of freedom, which more closely approximate the actual structure of the Universe, could potentially account for the features observed in the distance–redshift relation and may offer a better fit than models invoking phantom dark energy. 
Being this the case, further analyses considering alternatives approaches beyond the $\Lambda$LTB model will be crucial to robustly establish this finding. More complex spacetimes free of symmetrical assumptions~\cite{Celerier:2024dvs}, effective and phenomenological approximations based on backreactions effects with~\cite{Giani:2024nnv} and without cosmological constant~\cite{Lapi:2023plb,Lane:2023ndt,Seifert:2024bqr}, or generalized cosmographic approaches that do not rely on the assumption of a specific metric~\cite{Heinesen:2020bej,Maartens:2023tib,Adamek:2024hme} are some of the ways one could adopt to assess the effects of inhomogeneities at intermediate and small scales.

\section{Conclusions}\label{sec:conc}

The latest cosmological distance measurements have revealed late-time features in the distance-redshift relation that may be inconsistent with the expectations of $\Lambda$CDM. Adopting the $w_0w_a$ parameterization, the DESI collaboration has shown that a time-dependent dark energy component with a phantom equation of state can alleviate such discrepancies. While it is possible to develop physical models that mimic phantom behavior without violating the null-energy condition (see e.g. Ref.~\cite{Frion:2023xwq,Wolf:2025jed,Khoury:2025txd,Chen:2025mlf,Chen:2025wwn,Aoki:2025bmj,Lin:2025gne,deSouza:2025rhv,Dinda:2025iaq,Mirpoorian:2025rfp,Wang:2025zri,Kumar:2025etf,Murai:2025msx,Wolf:2025jed,Akrami:2025zlb,Pan:2025qwy,Chaussidon:2025npr,Silva:2025hxw,Chakraborty:2025syu,Li:2025dwz,Mishra:2025goj,Bhattacharjee:2025xeb,Braglia:2025gdo,Shiu:2025ycw,Bedroya:2025fwh,Philcox:2025faf,Giani:2025hhs}), it is crucial to investigate whether such late-time features could be explained by alternatives mechanism. Particularly, the observed distance-redshift relation could indicate deviations from the key assumption that distances are accurately described by the FLRW paradigm.  

Here, we focus on testing the large-scale inhomogeneity of the FLRW metric and its underlying assumption that density fluctuations do not impact the standard distance-redshift relation. We first leverage the recent DESI BAO measurements to test the geometric consistency of the $\Lambda$CDM model and $w_0w_a$ parametrization. This test relies on constraining the Hubble expansion history, considering the transverse and line-of-sight distances separately. Results of this test showed that different expansion rate histories are obtained when analyzing distances in these two different directions, hinting at an anisotropic expansion rate inconsistent with the FLRW metric. Such differences are particularly notable in the $\Omega_{\rm m}$ constraints. Overall, distances measured along the line of sight yield systematically lower values of $\Omega_{\rm m}$ compared to those measured in the transverse direction. While the statistical significance of these differences is, at most, modest, the trend poses a non-trivial challenge to the FLRW framework, as it persists regardless of the dark energy model considered.

Marking a distinctive feature, the observed trend further motivates the investigation of models that can explain distance measurements while naturally providing an anisotropic expansion rate. Adopting a simple late-time inhomogeneous generalization of $\Lambda$CDM, the $\Lambda$LTB model~\cite{Marra:2011ct,Valkenburg:2012td,Camarena:2021mjr}, we investigated whether cosmological data might reflect the effect of density fluctuations in the distance-redshift relation. Notably, our profile likelihood analysis reveals a systematic preference for $\Lambda$LTB scenarios featuring a spherical overdense region, whose size and amplitude depend on the data combination considered. Overall, the models allowed by the data provide an excellent fit—comparable, in most cases, to that achieved by the commonly adopted $w_0w_a$ framework. 

Our analysis shows that the combination of BAO distances and CMB priors allows for a broad parameter space, encompassing mildly overdense regions and extreme configurations in which homogeneity is restored only on Gpc scales. However, the most striking results emerge when supernova data are included. The addition of Pantheon+ supernovae, for example, significantly increases the preference for overdensities, challenging large-scale homogeneity at the $2.8\sigma$ level. The corresponding best-fit model describes a small but dense region ($r_{\rm B} \approx 130$ Mpc, $\delta_0 \approx 0.6$) and yields a substantially better fit than $\Lambda$CDM ($\Delta\chi^2_{\rm min} \approx -8.8$), statistically indistinguishable from the $w_0w_a$ parametrization. Importantly, this configuration---and most of the allowed parameter space in this case---remains consistent with the Copernican principle and can be interpreted as describing a scenario in which matter density fluctuations at intermediate and small scales give rise to the observed non-standard features in the distance–redshift relation. In contrast, the DESY5 supernova sample strongly favors more extended overdense regions, excluding the FLRW model at more than $5\sigma$. Its best-fit model features a broader but shallower overdensity ($r_{\rm B} \approx 800$ Mpc, $\delta_0 \approx 0.2$), with a fit comparable to that of the $w_0w_a$ model ($\Delta\chi^2 \approx -29$). Although this configuration is in $\sim10\sigma$ tension with the Copernican principle, DESY5 still allows more moderate scenarios that remain only mildly inconsistent with the assumption that we do not occupy a special place in the Universe.

While the assumption of large-scale homogeneity---and, consequently, the use of the FLRW metric to predict cosmological distances---has been foundational to our understanding of the Universe, the increasing precision of observational data may already be signaling a departure from this paradigm. We demonstrated that observed deviations from $\Lambda$CDM support an alternative interpretation which offers a statistically comparable fit to that of models invoking an evolving dark energy component with a phantom equation of state, although with a distinct and physically motivated origin: deviations from homogeneity at intermediate and small scales affect the distance-redshift relation. While this conclusion is based on a relatively simple inhomogeneous model, more sophisticated scenarios that accurately capture local density fluctuations could potentially offer a compelling explanation for the non-standard features observed in the data.

Raising concerns that the FLRW paradigm may fall short in predicting cosmological distances across a wide redshift range in the era of precision cosmology, our results motivate further investigations of scenarios that depart from the assumption of large-scale homogeneity. Moreover, as emphasized by various forecast analyses (see e.g. Ref~\cite{Euclid:2021frk,Euclid:2022ucc,Sakr:2023qpl}), upcoming surveys such as Euclid~\cite{EUCLID:2011zbd} and the Vera Rubin Observatory~\cite{LSSTScience:2009jmu}, which aim to improve the precision of cosmological distance measurements, will be crucial in determining whether these alternative scenarios can be robustly confirmed or ruled out. Additionally, the information contained in upcoming large-scale structure data could offer a crucial opportunity to test these scenarios, highlighting the importance of thoroughly investigating the signatures that inhomogeneities may imprint on large-scale observables~\cite{Marra:2022ixf}.

\begin{acknowledgments}
We are grateful to Leonardo Giani and Valerio Marra for comments on an early version of this manuscript and Adrià Gómez-Valent, Kyle Dawson, Andreu Font-Ribera, Collin Hill, Gordan Krnjaic, Oliver Philcox, Mustafa Amin, and Manoj Kaplinghat for useful discussion. This work was supported in part by the National Science Foundation grant OIA-2327192. D. C.~and F.-Y. C.-R.~would like to thank the Robert E.~Young Origins of the Universe Chair fund for its generous support. K.~G.~would like to thank the New Mexico (NM) Higher Education Department and the NM Space Grant Consortium for their support in this research. We also would like to thank the UNM Center for Advanced Research Computing, supported in part by the National Science Foundation, for providing the research computing resources used in this work.
\end{acknowledgments}

\appendix

\section{Methods} \label{ap:methods}

\subsection{Geometric analysis} \label{sub:ap_geo}

We test the geometrical consistency of the $\Lambda$CDM and $w_0w_a$CDM models (see Section~\ref{sub:geo_test}) by performing a Markov Chain Monte Carlo (MCMC) analysis of their parameter space. 
Particularly we employ \texttt{MontePython}~\cite{Brinckmann:2018cvx, Audren:2012wb} to perform the statistical analysis and \texttt{CLASS}~\cite{CLASS} to compute the theoretical predictions of both models. 
The convergence of the MCMC chains is assessed via Gelman–Rubin criterion~\cite{10.1214/ss/1177011136}, represented by $R$, requiring $R - 1 \lesssim 0.01$ for all cases.

In the case of $\Lambda$CDM, when analyzing only cosmological distances, we restrict the parameter space to two variables: $r_{\rm d} h$ and $\Omega_{\rm m}$, where $h$ is the dimensionless Hubble parameter defined via $H_0 = 100\,h\,\,\mathrm{km\,s^{-1}\,Mpc^{-1}}$. This choice reflects the fact that uncalibrated SNeIa and BAO data cannot independently constrain $H_0$ and $r_{\rm d}$. This approach requires a minor modification to the theoretical predictions from \texttt{CLASS}, which by default computes cosmological distances as functions of $\Omega_{\rm m}$ and $h$, i.e., $D_x \equiv D_x(z; \Omega_{\rm m}, h)$, where $D_x$ could represent $D_{\rm M}$, $D_{\rm H}$, or $D_{\rm L}$.

When analyzing BAO data, which constrain the ratios $D_{\rm M}/r_d$ and $D_{\rm H}/r_d$, we address the degeneracy between $r_d$ and $h$ by adopting normalized distances defined as $\hat{D}_x(z; \Omega_{\rm m}) \equiv h\,D_x(z; \Omega_{\rm m}, h)$. This allows us to express the observables as
\begin{equation}
    \frac{D_x(\Omega_{\rm m}, h)}{r_{\rm d}} = \frac{\hat{D}_x(\Omega_{\rm m})}{r_{\rm d} h}\,.
\end{equation}
A similar strategy is used in the analysis of uncalibrated SNeIa distances, which can be also assessed using $\hat{D}_\mathrm{L}$. This is justified by the fact that the distance modulus $\mu(z)$ yields strong degeneracy between $h$ and $M$, the absolute magnitude of SNeIa. Since $M$ is treated as a nuisance parameter and marginalized over in the final analysis, the use of normalized distances has no impact on the inferred cosmological parameters.

The normalization procedure becomes unnecessary when including the CMB priors on $(\theta_*, \omega_{\rm b}, \omega_{\rm c})$. In that case, we directly sample the parameter space in terms of $h$, $\omega_{\rm b}$, and $\omega_{\rm c}$. For the $w_0w_a$CDM parametrization, we follow analogous approaches, depending on whether the analysis considers the CMB priors, and extending the parameter space to include the dark energy equation-of-state parameters $w_0$ and $w_a$.

\subsection{Best-fits and profile likelihood analysis}\label{sub:profile_like}

We use a tailored version of the \texttt{voiddistance2020}~\cite{Valkenburg:2012td} code to compute the theoretical predictions of the $\Lambda$LTB model.\footnote{Our implementation is available at \url{https://github.com/davidcato/monteLLTB}.} We determine the global best-fit models and corresponding profile likelihoods by minimizing the relevant likelihood functions using the {Derivative-Free Optimizer for Least-Squares} (DFOLS)~\cite{10.1145/3338517} and the {Covariance Matrix Adaptation Evolution Strategy} (CMA-ES)~\cite{hansen2019pycma} packages. Our strategy begins with CMA-ES, which efficiently explores the parameter space and provide a preliminary best-fit model without requiring an accurate initial guess. This result is then refined using DFOLS, which allows for efficient local optimization around the preliminary solution at low computational cost. To unsure the underlying best fit does not correspond to a local minimum, we repeat this process iteratively until a stable solution is found. 

We profile the likelihood over thirty-one different values of $\delta_0$ for the one-dimensional case, and over a $(16,18)$ grid in the $\delta_0$--$z_{\rm B}$ parameter space for the two-dimensional profiling. The $\delta_0$ values are chosen to be homogeneously distributed within the range $\delta_0 \in [-0.2,\, 0.625]$, while the $z_{\rm B}$ values follow an inhomogeneous spacing within the range $z_{\rm B} \in [0,\, 1.5]$. 
The values obtained in the corresponding parameter space are then interpolated to produce Fig.~\ref{fig:chi2_and_LLTB_contours}; a third-order spline interpolation is used for the one-dimensional case, while a fourth-order spline is employed for the two-dimensional cases.

We determine the values of $\Delta \chi^2_{\rm 2d} \equiv \chi^2(\delta_0, z_{\rm B}) - \chi^2_{\rm min}$ corresponding to the 68\% and 95\% percentiles in the two-dimensional analysis using the cumulative distribution function ${\rm CDF}(\Delta \chi^2_{\rm 2d})$. This function is reconstructed by assigning a weight to each point in the $\delta_0$--$z_{\rm B}$ grid, given by $\exp\left(-\frac{1}{2} \Delta \chi^2_{\rm 2d}\right)$.

To assess the statistical significance of the $\Lambda$LTB model for a given dataset, we use the cumulative distribution function from the one-dimensional analysis, ${\rm CDF}(\Delta \chi^2_{\rm 1d})$. We adopt the one-dimensional case because fixing $\delta_0 = 0$ is sufficient to recover the FLRW paradigm. Following the DESI analysis~\cite{DESI:2025zgx}, we define:
\begin{equation}
    {\rm CDF}(\Delta\chi^2_{\rm 1d}) = \frac{1}{\sqrt{2\pi}} \int_{-N}^{+N} e^{-t^2/2}\; {\rm d}t\,,
\end{equation}
which corresponds to the standard Gaussian interpretation of an $N\sigma$ confidence level for a one-dimensional normal distribution. Alternatively, one could adopt ${\rm CDF}(\Delta \chi^2_{\rm 2d})$ to evaluate the consistency of the FLRW framework at $\delta_0 = 0$ and $z_{\rm B} = 0$. We find that this approach yields slightly stronger significance in favor of $\Lambda$LTB models.

\subsection{Supernova binning}\label{sub:SNe_binning}

Figure~\ref{fig:6_panel} shows the binned version of DESY5 and Pantheon+ data points. To produce such points, we compute $\Delta \mu \equiv \mu - \mu_{\textrm{fid}}$ for each supernova, where $\mu_{\textrm{fid}}$ is the distance modulus predicted by the $\Lambda$CDM fiducial model used in the DESI BAO analysis~\cite{DESI:2024cru}. These values, along with their corresponding redshifts, were then binned using the procedure described in~\cite{Schmelling_1995}, with 10 bins defined to be equally spaced in logarithmic redshift over the relevant redshift range. 

\section{Extended Geometrical Test Results}

\subsection{Features in the $w_0$--$w_a$ space \label{ap:geo_w0wa}}

The $w_0w_a$ parametrization features strong correlations that make difficult to robustly interpret its constraints. To partially overcome this issue, we present our results using the value of the dark energy EoS at the pivot redshift, $w_\mathrm{pivot}$, instead of $w_0$. As discussed in Ref.~\cite{Cortes:2024lgw}, the main advantage of using $w_\mathrm{pivot}$ is that it decorrelates from $w_a$ and provides a more direct measure of the EoS at the effective redshift where the data combination maximizes its statistical power. The main limitation, however, is that the value of the pivot redshift $z_\mathrm{pivot}$ depends on the specific dataset, which complicates direct comparisons across different analyses. To avoid this complication, we fix $z_\mathrm{pivot} = 0.32$ for all cases, since this is the value that minimizes the correlation between $w_\mathrm{pivot}$ and $w_a$ in the cases of interest.

\begin{figure}
    \centering
    \includegraphics[width=0.975\linewidth]{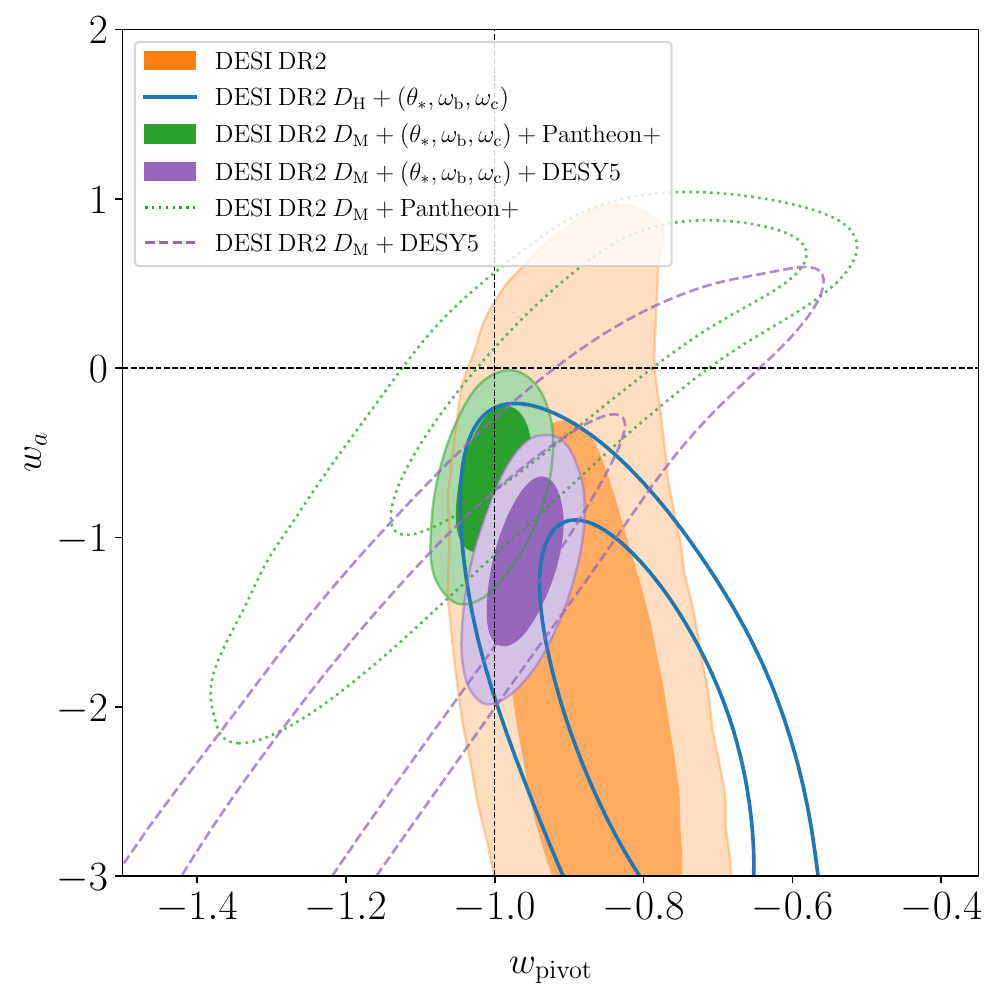}
    \caption{Marginalized constraints in $w_\mathrm{pivot}$ and $w_a$ at $68\%$ and $95\%$ credible intervals obtained from the geometrical test $D_\mathrm{M}$ vs. $D_\mathrm{H}$.}
    \label{fig:w0wa}
\end{figure}

Figure~\ref{fig:w0wa} shows the results of the geometrical test in the $w_\mathrm{pivot}$–$w_a$ parameter space. 
As anticipated, the impact of changes in the expansion rate is less pronounced in these parameters, primarily due to limited statistical power and the fact that cosmological distances only probe indirectly the evolution of the dark energy equation of state \cite{Wang:2004ru,Wang:2025vfb}.
Nevertheless, interesting features emerge. Analyses based on $D_\mathrm{M}$---whether or not a CMB prior is included (purple and green contours and lines)---are highly consistent with $w_\mathrm{pivot} = -1$. In contrast, constraints using $D_\mathrm{H}$ measurements (orange contours and blue lines) exhibit a slight preference for larger values of the EoS at $z_\mathrm{pivot}$. The preference for an evolving EoS, i.e., $w_a \neq 0$, also varies across the different cases shown. In particular, uncalibrated transverse distances from BAO and Pantheon+ (dotted green line) are well described by a model with $w_a = 0$; however, this model shows mild tension with the data once the CMB prior is taken into account (solid green). Including DESY5 supernovae distances (purple lines and contours) reduces the preference for a cosmological constant even further. Similarly, expansion rates inferred from line-of-sight distances and the CMB prior exclude the $w_a = 0$ scenario at more than $2\sigma$.

\subsection{Constraints from Pantheon$+$ data \label{ap:geo_Pantheon}}

\begin{figure}
    \centering
    \includegraphics[width=0.975\linewidth]{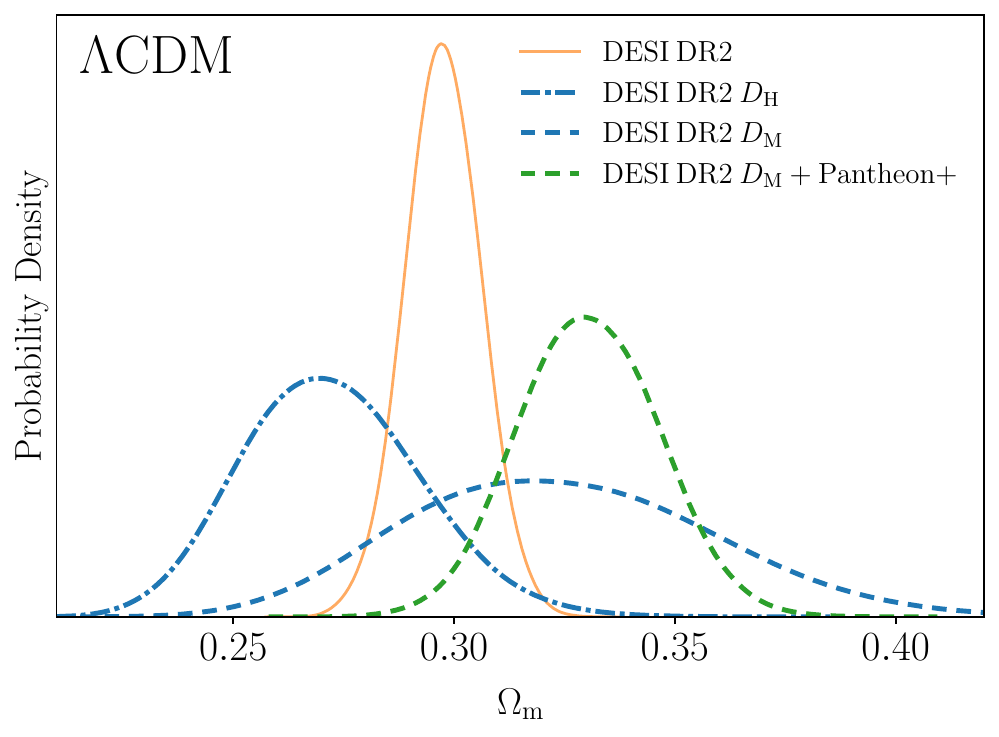}
    \includegraphics[width=0.975\linewidth]{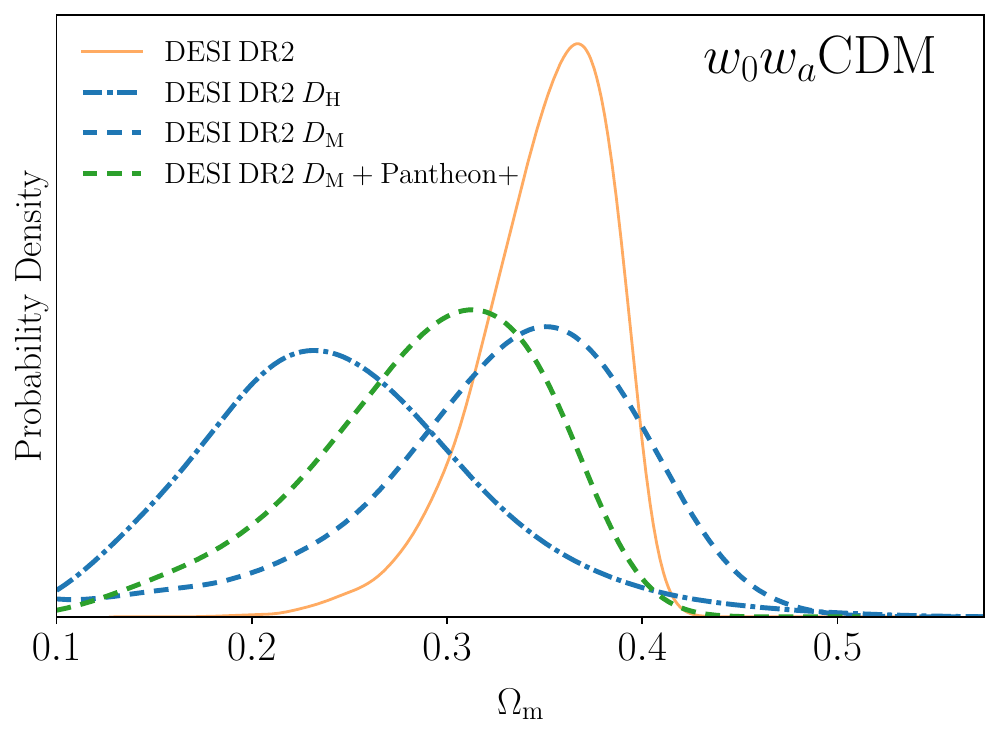}
    \caption{Marginalized $\Omega_\mathrm{m}$ posterior obtained from the geometrical test, $D_\mathrm{M}$ vs. $D_\mathrm{H}$, when employing Pantheon$+$ data and considering both the $\Lambda$CDM model (upper panel) and $w_0w_a$ parametrization (lower panel).}
    \label{fig:Omega_m_Pantheon}
\end{figure}

Figure~\ref{fig:Omega_m_Pantheon} shows the marginalized constraints on $\Omega_{\rm m}$ from the geometrical test using Pantheon$+$ data, under $\Lambda$CDM (upper panel) and $w_0w_a$CDM (lower panel). As pointed out in Section~\ref{sub:geo_test}, combining Pantheon$+$ SNeIa with BAO transverse distances yields higher values of $\Omega_{\rm m}$ compared to constraints from direct measurements of the expansion rate via $D_\mathrm{H}$. However, relative to the analysis with DESY5 data, the shift in $\Omega_{\rm m}$ is less pronounced.

\section{LTB (extra) degrees of freedom} \label{ap:LTB_extra}

The arbitrary functions of the LTB metric emerge as constants of integration when solving the Einstein and generalized Friedmann equations. The first of these is the curvature function $k(r)$, which generalizes the constant spatial curvature of FLRW models. It arises from one of the Einstein equations, which yields (see, e.g.,~\cite{Enqvist:2007vb}):
\begin{equation} \label{eq:kr}
    X(r,t) = \frac{\partial_r A(r,t)}{\sqrt{1 - k(r)\,r^2}}\,.
\end{equation}

A second equation, analogous to the FLRW acceleration equation,
\begin{equation*}
    \partial_t A^2(r,t) + 2 A(r,t) \partial^2_t A(r,t) + k(r) r^2 = \Lambda\, A^2(r,t)\,,
\end{equation*}
leads to a generalized Friedmann equation:
\begin{equation} \label{eq:friedmann_LTB}
    H^2_{\perp}(r,t) = \frac{\Lambda}{3} + \frac{2m(r)}{A^3(r,t)} - \frac{k(r) r^2}{A^2(r,t)}\,,
\end{equation}
where $m(r)$ is the so-called mass function, representing the second arbitrary degree of freedom. This function is directly related to the matter energy density via
\begin{equation*}
    4\pi G \rho_{\rm m}(r,t) = \frac{\partial_r m(r)}{\partial_r A(r,t)\, A^2(r,t)}\,.
\end{equation*}

The third arbitrary function is the Big Bang time function, $t_{\rm BB}(r)$, which arises when computing the age of the Universe, $t_0$. It defines the surface of the initial singularity and is given by
\begin{equation} \label{eq:t0_LLTB}
    t_0 - t_{\rm BB}(r) = \int^1_0 \left[\frac{2m(r)}{A_0^3 x^3} - \frac{k(r)\,r^2}{A_0^2 x^2} + \frac{\Lambda}{3} \right]^{-1/2} \frac{dx}{x}\,,
\end{equation}
where we have used $A_0 \equiv A(r,t_0)$.

Eq.~\eqref{eq:t0_LLTB} shows that the three functions introduced above are not entirely independent, as one can be expressed in terms of the others. Additionally, the spherical symmetry of the $\Lambda$LTB model introduces a coordinate gauge freedom related to the definition of the radial coordinate. Here, we fix this freedom by choosing the mass function to scale as $m(r) \propto m_0 r^3$, where $m_0$ is an arbitrary mass scale.

Moreover, since we focus on late-time effects, we restrict our $\Lambda$LTB model to a subclass of inhomogeneous models that closely recover the FLRW metric at early times and leave the standard inflationary paradigm unaltered. These ``early-FLRW cosmologies''~\cite{Marra:2022ixf} can be realized by setting $t_{\rm BB}(r) = 0$ and are, in general, expected to minimally affect the CMB.\footnote{Allowing $t_{\rm BB}(r) \neq 0$ would otherwise introduce decaying modes, yielding inconsistencies with standard inflation~\cite{Silk:1977aaa,Zibin:2008vj}.}

With these choices, the only remaining free function is the curvature profile $k(r)$, which can be specified according to the physical scenario under study. Since we use the $\Lambda$LTB model as a diagnostic tool to investigate how subtle deviations from the large-scale inhomogeneity could alleviate cosmological distance discrepancies, we adopt a monotonic compensated curvature profile given by
\begin{equation}\label{eq:kr}
    k(r) = k_{\rm B} + (k_{\rm C} - k_{\rm B})\, P_3(r/r_{\rm B})\,,
\end{equation}
where $k_{\rm B}$ and $k_{\rm C}$ are the values of the curvature at the boundary and at the center of the inhomogeneity, respectively. Likewise, $r_{\rm B}$ denotes the comoving boundary radius of the inhomogeneity in radial metric coordinates. The function $P_n(x)$ encodes the radial profile of the matter distribution and is defined as~\citep{Valkenburg:2012td}:
\begin{align}\label{eq:kr_P}
P_n(x) = 
\begin{cases}
1 - \exp\left[-\dfrac{(1 - x)^n}{x}\right] & \text{for } 0 \leq x < 1 \,,\\
0 & \text{for } x \geq 1\,.
\end{cases}
\end{align}

This profile ensures that the FLRW metric is recovered at scales $r \geq r_{\rm B}$. Furthermore, it implies the existence of a compensating radius $r_L < r_{\rm B}$ at which the matter density contrast,
\begin{equation}\label{eq:delta_rho}
    \delta_{\rm m}(r,t) = \frac{\rho_{\rm m}(r,t)}{\rho_{\rm m}(r_{\rm B},t)} - 1\,,
\end{equation}
changes sign.

We assume an asymptotically flat FLRW geometry by setting $k_{\rm B} = 0$. Therefore, the additional degrees of freedom induced by the LTB finally reduce to two free parameters: the central curvature $k_{\rm C}$ and the size of the inhomogeneity $r_{\rm B}$. This parameter space can be alternatively recasted in terms of more physically intuitive quantities: the present-day matter density contrast at the center, $\delta_0 \equiv \delta_{\rm m}(0, t_0)$, and the redshift $z_B$ at which the FLRW metric is effectively recovered.

\section{CMB observables in the $\Lambda$LTB model} \label{ap:inho_CMB}

One of the key challenges in inhomogeneous cosmologies is the consistent assessment of CMB observables, which depends strongly on both the observer's position and the specific details of the model. For instance, an off-center observer perceives the CMB as anisotropic, typically dominated by a prominent dipole component arising from asymmetric gravitational redshifts~\cite{Alnes:2006pf,Moffat:2005yx,Foreman:2010uj}. Moreover, for observers falling toward the center of the overdensity, this gravitationally induced dipole can align with the intrinsic dipole due to peculiar motion, further complicating the interpretation of CMB anisotropies~\cite{Rakic:2006tp}. The presence of inhomogeneities at early times, on the other hand, can lead to a breakdown of the standard cosmological perturbation framework commonly adopted for analyzing CMB phenomena. In such cases, it becomes necessary to study perturbations on an inhomogeneous background metric to enable meaningful comparisons with CMB observations (see, e.g., Refs.~\cite{Clarkson:2007yp,Zibin:2008vj}).

Nonetheless, the $\Lambda$LTB model employed here avoids these complexities by construction: it is designed to recover the FLRW framework at early times, and the observer is assumed to reside very close to the center of the inhomogeneity. Thus, under the assumption of a fixed primordial curvature power spectrum, density fluctuations modify the CMB spectrum solely through line-of-sight effects—namely, the late-time Integrated Sachs-Wolfe (ISW) effect and the angular diameter distance to the last scattering surface. Depending on the exact configuration of the inhomogeneity, these effects can lead to significant modifications or, conversely, may simply reproduce an ISW signal consistent with predictions from linear perturbation theory in an FLRW background~\cite{Valkenburg:2012td}. In the latter case, the CMB spectra can be accurately computed using an effective FLRW model that encodes the line-of-sight effects induced by late-time inhomogeneities~\cite{Zibin:2008vk,Marra:2010pg,Biswas:2010xm}. Comparison with observed power spectra constrains the cosmological parameters of this effective $\Lambda$CDM model, thereby providing indirect constraints on the parameters describing the underlying inhomogeneous model (see, e.g., Ref.~\cite{Camarena:2021mjr}). The assumption of a compensated profile further facilitates the construction of such effective models, as it establishes a finite scale at which the homogeneous limit is recovered and allows for a direct comparison between FLRW and LTB distances.

Although this approach is suitable for analyzing the CMB power spectrum, in this work we adopt a multivariate Gaussian prior on $(\theta_*, \omega_{\rm b}, \omega_{\rm c})$~\cite{Lemos:2023xhs}. As previously discussed, such a prior compresses the CMB data to extract information that remains largely insensitive to late-time effects, such as the Integrated Sachs-Wolfe (ISW) signal. Given the limited sensitivity to late-time phenomena, we impose the CMB prior on the $\Lambda$LTB cosmological parameters that describe the background outside the inhomogeneous region—except for the angular acoustic scale, which is instead computed as the CMB angular scale measured by the effective FLRW observer discussed earlier. Yet, further analyses show that applying the $\theta_*$ prior either to the background $\Lambda$LTB model or to the effective $\Lambda$CDM model yields effectively the same results, reinforcing the fact that the class of models considered here only marginally affects CMB constraints.

\bibliography{distances}

\end{document}